Parker Solar Probe evidence for the absence of whistlers close to the Sun to scatter strahl and regulate heat flux

C. Cattell[1], A. Breneman[12], J. Dombeck[1], E. Hanson[1], M. Johnson[1], J. Halekas[2], S. D. Bale[3,10], T. Dudok de Wit[7], K. Goetz[1], K. Goodrich[11], D. Malaspina[8,9], M. Pulupa[3], T. Case[6], J. C. Kasper[4,5], D. Larson[3], M. Stevens[6], P. Whittlesey[3]

1. School of Physics and Astronomy, University of Minnesota, 116 Church St. SE Minneapolis -mail:cattell@umn.edu

2. Department of Physics and Astronomy, University of Iowa, Iowa City, IA 52242, USA

3. Space Sciences Laboratory, University of California, Berkeley, CA 94720-7450, USA

4. BWX Technologies, Inc., Washington DC 20002, USA

5. Climate and Space Sciences and Engineering, University of Michigan, Ann Arbor, MI 48109, USA

6. Smithsonian Astrophysical Observatory, Cambridge, MA 02138 USA

7. LPC2E, CNRS, CNES and University of Orléans, Orléans, France

8. Laboratory for Atmospheric and Space Physics, University of Colorado, Boulder, CO 80303, USA

9. Astrophysical and Planetary Sciences Department, University of Colorado, Boulder, CO, USA

10. Department of Physics, University of California, Berkeley, Berkeley, CA 94708

11. Department of Physics and Astronomy, West Virginia University, Morgantown, WV 26506-6315

12. NASA Goddard Space Flight Center, 8800 Greenbelt Rd, Greenbelt, MD 20771

Abstract: Using the Parker Solar Probe FIELDS bandpass filter data and SWEAP electron data from Encounters 1 through 9, we show statistical properties of narrowband whistlers from ~16 $R_s$ to ~130 $R_s$, and compare wave occurrence to electron properties including beta, temperature anisotropy and heat flux. Whistlers are very rarely observed inside ~28 $R_s$ (~0.13 au). Outside 28 $R_s$, they occur within a narrow range of parallel electron beta from ~1 to 10, and with a beta-heat flux occurrence consistent with the whistler heat flux fan instability. Because electron distributions inside ~30 $R_s$ display signatures of the ambipolar electric field, the lack of whistlers suggests that the modification of the electron distribution function associated with the ambipolar electric field or changes in other plasma properties must result in lower instability limits for the other modes (including the observed solitary waves, ion acoustic waves) that are observed close to the Sun. The lack of narrowband whistler-mode waves close to the Sun and in regions of either low (<.1) or high (>10) beta is also significant for the understanding and modeling of the evolution of flare-accelerated electrons, and the regulation of heat flux in astrophysical settings including other stellar winds, the interstellar medium, accretion disks, and the intra-galaxy cluster medium.

1. Introduction

The role of whistler-mode waves in the evolution of solar wind electrons has been of interest since the prediction of changes in electron distributions with distance from the Sun by Feldman et al. (1975) and the early Helios-1 wave observations of Gurnett and Anderson (1977) and Neubauer et al. (1977) near 1 AU. Proposed instability mechanisms included temperature anisotropy (Kennel and Petscheck, 1966; Gary and Wang, 1996) and heat flux (Forslund, 1970; Feldman et al., 1975; Gary, 1978). Most research has focused on the role of whistlers in regulating electron heat flux and scattering of strahl to produce halo electrons. As spacecraft data at a wide range of radial distances, as well as out of the ecliptic plane, variations with radial distance and latitude could be studied. Changes in the occurrence and properties of whistler-mode waves have been documented from ~.3 au to 1 au (Beinroth and Neubauer, 1981; Jagarlamudi et al., 2020). Lengyel-Frey et al. (1996) compared Ulysses (~1 to 5 au) and Helios (~.3 au to 1 au) observations of whistlers; Lin et al. (1998) used Ulysses data to characterize waves both in and out of the ecliptic. Radial variation in the electron properties was shown using Helios from .3 au to 1 au (Maksimovic et al., 2005), using Helios, Cluster and Ulysses out to ~4 au (Štverák et al., 2009) and, with Cassini, out to >5 au (Graham et al., 2017). At 1 au, Bale et al. (2013) found that there was a transition in heat flux regulation from collisional to collision-less regimes that depended on the Knudsen number. Scaling of the heat flux depended on beta in the collision-less regime. Scime et al. (1994) showed that, from ~1 to 5 au in the ecliptic plane, the changes in the electron heat flux were consistent with whistler scattering.

Recent studies of waveform data from STEREO at 1 AU (Cattell et al., 2020) and Parker Solar Probe (Agapitov et al., 2020; Cattell et al., 2021a) have provided evidence that the whistler heat flux fan instability (Bošková et al., 1992; Krafft and Volokitin, 2003; Vasko et al., 2019), is consistent with simultaneous wave and electron observations. Electron distributions are consistent with scattering by the whistlers (Cattell et al., 2021b). A larger study of the electron heat flux from Encounters 1 through 5 outside ~28 Rs (~.13 au) by Halekas et al., (2020) found that heat flux measurements were bounded by this instability limit and the oblique fast mode whistler instability (Verscharen et al. (2019). Simulations, including both particle-in-cell (Micera et al., 2020, 2021; Roberg-Clark et al, 2018, 2019) and particle tracing (Cattell and Vo, 2021; Vo et al., 2021), have indicated that, in addition to oblique waves, parallel or quasi-parallel whistlers may also play a role.

Initial studies from Parker Solar Probe's encounters 1 through 4 using waveform capture data found that whistler-mode waves occurred less often inside ~35 Rs (Cattell et al. 2021a). However, the occurrence was not normalized by the number of waveform captures, and, in addition, because the largest captures are stored and transmitted, these data do not provide a statistical sample useful for determining occurrence rate. Similarly, although STEREO waveform captures provided evidence for large amplitude oblique whistlers at ~1 au and showed that they occurred with a large packing fraction in stream interactions regions and coronal mass ejections, occurrence rates could not be determined.

Using ARTEMIS data near 1 au, Tong et al. (2019) found small amplitude quasi-parallel whistlers in ~2% of the wave spectra samples in the pristine solar wind. Although the rate increased with electron temperature anisotropy, the waves were consistent with the heat flux instability. Similar wave properties were observed in Cluster data in the slow solar wind (Lacombe et al., 2014). A statistical study using the PSP BPF magnetic field data during Encounter 1 at ~.2 au (Jagarlamudi et al 2021) found a whistler rate of ~1.5%, and that wave occurrence was consistent with the heat flux fan instability.

In this study, we focus on changes with distance from the Sun in the occurrence probability of whistler-mode waves and the association with electron properties during Parker Solar Probe Encounters 1 through 9 (E1 to-E9) where 'encounter refers to the portion of the orbit associated with a given periapsis when data were obtained.' The data sets and methodology are described in section 2, statistical results are presented in section 3, and discussion and conclusions are given in section 4.

2. Data sets, methodology

We utilize Level 2 data from the Parker Solar Probe (PSP) FIELDS (Bale et al., 2016) instrument suite and the Level 2 and 3 data from SWEAP Instrument suite (Kasper et al., 2016. The primary FIELDS data set is the DC coupled bandpass filter (BPF) data, usually obtained at a cadence of 1 spectrum every 0.87 s. The electric field covers a frequency range from 0.4 Hz to 8000 Hz and the search coil magnetic field channel covers a frequency range from 4 Hz to 8000

Hz. The electric field component was V34; and the magnetic field component varied (SCMU in Encounter 1(E1) and E2, and SCMV in E3-E9). Higher frequency resolution is provided by the AC and DC coupled spectral data, which are usually obtained at a rate of 1 spectra every ~28 seconds, over a frequency range of ~10 Hz to 4.8 kHz (Malaspina et al., 2016). For Encounter 1, we also utilized the dc coupled cross-spectral data to obtain the wave propagation angles. Electric field amplitudes were determined from measured voltages assuming the physical boom length of 3.5 meters. The quasi-static magnetic field data in RTN (radial-tangential-normal) coordinates at ~4 samples per second are used to determine the electron cyclotron and lower hybrid frequencies.

From the SWEAP Instrument suite (Kasper et al., 2016), we show the pitch angle distribution for the 314 eV channel, indicative of strahl, and utilize electron heat flux, parallel electron beta ($\beta_{e,\|}$), and core temperatures (Halekas et al., 2020, 2021) to compare to instability mechanisms. These parameters are obtained using Level 3 moments from the Solar Probe Cup (SPC) (Case et al., 2020), ion data from SPAN-I, and L2 electron data from the Solar Probe Analyzers (SPAN-A-E and SPAN-B-E)(Whittlesey et al. 2020). Note that no electron data for Encounter 3 are used because the accurate solar wind velocity measurements needed to transform the electron distributions into the plasma frame were not available.

To identify the narrowband whistler-mode waves in the BPF data, we employ an approach based on one developed for Van Allen Probes (Tyler et al. 2019a, 2019b). Because the PSP BPF data set has 13 channels, whereas the Van Allen Probes data had 7 channels (only occasionally using 13), a number of modifications were implemented. We developed an auto-identification code using a non-linear fitting approach to identify BPF samples consistent with a narrowband wave and to determine their frequency and amplitude, by minimizing the error between the measured BPF values and the expected response of the BPF to a sine wave with this frequency and amplitude. Details of the code are given in the Appendix. A BPF sample is identified as containing a narrowband whistler-mode wave when a peak in observed in both the electric field and the magnetic field at the same frequency (within 25%), within the band between the lower hybrid frequency, $f_{lh}$, and 0.5 $f_{ce}$, and with an error <0.1. In practice, in approximately half of the events, the frequencies identified in the electric and magnetic fields were the same within 4%, and almost all identified waves had frequencies $\leq .2 f_{ce}$ (less than .1% were between 0.2 and 0.5 $f_{ce}$).

An indication of the accuracy of the auto-id code is provided in Figure 1, which compares the times identified as containing whistlers with the original BPF data during a 5 minute interval from E1. The white lines overplotted on the BPF plots and the spectrum are 0.2 $f_{ce}$ and $f_{lh}$. It is clear that the code identifies almost all the same times as one identifies by eye; only a few low amplitudes wave intervals were not identified. The improved frequency resolution provided by the code is also apparent. Comparing the fast time resolution of the BPF (panel d) and the better frequency resolution of the spectral data (panel e) shows that the code very accurately tracks the peak frequency at the higher time resolution. Panels a, b and c show that the wave amplitudes and propagation angles can vary rapidly, There was one waveform capture in the interval (at 10:43:47, shown in Figure 1 of Cattell et al., 2021b). Note that comparisons to waveform capture data have been made for intervals in several different encounters, and these data also have very sinusoidal waveforms and show the right-hand circular or elliptical polarization expected for whistlers, providing additional confirmation that the auto-id code is correctly finding the narrowband whistlers. Note that the amplitudes identified by the code are a lower bound since they are based on a single component.

## 3. Observations and statistics

Encounter 7 (exhibits several of the features of interest, as shown in Figure 2 as shown in Figure 2, for 9 through 25 Jan, 2021. The top three panels plot information from the auto-id code: the PSP orbit versus radial distance in AU with the coverage of the BPF data as orange dots and identified whistlers as green dots; the wave electric field amplitudes for identified whistlers; and the wave frequency in green; the electron cyclotron frequency, $f_{ce}$, in blue, 0.3 $f_{ce}$ in yellow, and the lower hybrid frequency, $f_{lh}$, in red are also plotted. The next panels plot the magnetic field in RTN coordinates, the proton flow speed from SPAN-I, and DC-coupled spectra of the magnetic field and the electric field from 30-5000 Hz, electron energy spectrum and 314 eV pitch angle spectrum. Two features are immediately obvious. There are many more whistlers on the outbound pass than on the inbound one. This asymmetry is consistent with earlier observations showing that whistlers are more frequently observed in the slow solar wind during E1, ~.2 au (Jagarlamudi et al., 2021), in Helios data from .3 to 1 au (Jagarlamudi et al., 2020), and at 1 au (Lacombe et al., 2014). There are almost no whistlers near periapsis; instead large amplitude electrostatic waves occur. The broadband electrostatic waves at closest approach include nonlinear electrostatic waves and solitary waves (also referred to as time domain structures), previously identified in E3 and E5 (Mozer et al., 2021a). There are two regions of narrowband electrostatic waves at slightly higher radial distances ~23 to 28 Rs, during the inbound orbit on 16 January, and the outbound orbit on 18 and 19 January. Waves on the outbound orbit have been identified as oblique ion acoustic waves (Mozer et al. 2021b). Whistlers are not observed inside ~28 Rs on the outbound orbit and ~32 Rs on the inbound orbit. Wave amplitudes are up to ~3**5** mV/m on the outbound orbit, and up to ~22 mV/m on the inbound. The plasma density is larger (~500 to ~1500/ $cm^3$) and the solar wind speed slower (~200 km/s) in the outbound region with whistlers compared to <200/$cm^3$ and ~400 to ~600 km/s during the inbound intervals with whistlers.

Figures 3 and 4 summarize the statistical results. The total number of BPF samples versus radial distance is plotted in 4a, the number of samples identified as whistlers in 3b, and 3c shows the rate of whistler occurrence. The red line at ~28 Rs (~0.13 au) is to guide the eye to show that there are significant numbers of BPF samples inside this radial distance, but very few contain whistlers. It is clear that the whistler occurrence rate drops off dramatically inside ~28 Rs.

Figure 3d shows the wave frequency identified in the electric field, and 3e shows the ratio of the spacecraft frame frequency to the electron cyclotron frequency, f/fce. The wave frequency decreases with radial distance; f/fce is usually less than 0.2, but increases with radial distance. The wave electric field amplitude, dE, is shown in panel f, the wave magnetic field amplitude, dB, in panel h, and the ratio of wave magnetic field to background magnetic field, dB/B, in panel i. The electric and magnetic field amplitudes decrease with radial distance; dB/B is approximately constant. It is interesting that the few whistlers seen inside 28 Rs are very large amplitude. The mean frequency at 30 Rs (70 Rs) is 164 Hz(76 Hz); the mean electric field amplitude is 20 mV/m(6 mV/m); and the wave magnetic field relative to the background magnetic field is .009(.013). These results are consistent with the study of waveform capture data from E1–E4 (Cattell et al., 2021a) in the shared radial distances (outside ~30 Rs). Panel **3g,** which plots the ratio of the electric field amplitude (34 component) to the component of the magnetic field (SCMU in E1 and E2, and SCMV in E3–E9) provides an estimate of the wave phase velocity. This is only a rough estimate because the orientation of the spacecraft with respect to the solar wind magnetic field is highly variable, and only a single component of each field is used. This ratio varies from ~7500 km/s inside ~50 Rs to ~3000 km/s outside ~110 Rs.

The decrease in whistler occurrence seen at r ≳ 75 Rs (~.3 au) is very likely instrumental. Wave frequencies are decreasing with radial distance as shown in Figure 4d. At 1 au, the median wave frequency for narrowband whistlers was 38 Hz (Cattell et al., 2020). Because the auto-id algorithm uses two bins on either side of the bin with the maximum amplitude to determine the wave frequency, no whistlers can be identified that have frequencies in the lowest two bins. In practice, this means only waves with frequencies above ~50 Hz can be accurately identified using this algorithm.

The relationship between selected electron parameters and the occurrence of whistlers is given in Figure 4. The left-hand plots show measurements obtained when there were no whistler waves identified within 15 seconds, and the right hand plots are for measurements when there was at least one whistler identified within 15 seconds. 4a and 4b, the normalized heat flux, $Q_{norm}$, and 4c and 4d, $\beta_{e,||}$, are plotted versus radial distance and color coded by encounter number. 4e through 4h are plotted versus $\beta_{e,||}$, and color coded by radial distance. Note that 4f and 4h are for a restricted range of $\beta_{e,||}$. Both 4a and 4b show that the $Q_{norm}$ decreases significantly closer to the Sun. Although both the measured heat flux and the saturation heat flux are increasing, the increase in the saturation heat flux is larger. 4b also shows the decrease in whistler occurrence inside ~28 Rs. The excursions to very large beta seen in 4c are associated with heliospheric current sheet crossings. 4d reveals that whistler waves are seen only in a limited range of $\beta_{e,||}$, from ~1 to ~10, as was also seen for the oblique whistlers observed in the STEREO waveform data at 1 au (Cattell et al., 2020). Panels e and f plot $Q_{norm}$ versus $\beta_{e,||}$ with the heat flux fan instability threshold from Vasko et al (2019) over-plotted. Panel f shows that the wave occurrence is constrained by this threshold. Note that wave occurrence is also constrained by the Lacombe et al (2014) and Gary et al (1999) (parallel) heat flux instability limits (not shown). There are two interesting features in measurements obtained when no whistlers were observed (5e). At low beta (<~.1), $Q_{norm}$ is well below instability thresholds and these observations were almost all obtained inside ~25 Rs. At very high beta (>50), which was observed over the range of radial distances, $Q_{norm}$ values are well above the threshold, with no dependence on beta, indicating that neither the whistler heat flux fan instability nor the (parallel) heat flux instability are operating.

Panels 4g and 4h plot the core electron temperature anisotropy $T_{e\perp}/T_{e\parallel}$ vs $\beta_{e,\parallel}$ with the temperature anisotropy and firehose instability limits over-plotted. The upper dotted line is the whistler temperature anisotropy threshold, and the lower lines are arbitrary firehose resonant (dot-dash) and non-resonant (dash) instability (both from Lacombe et al. 2014, based on Gary and Karimabadi, 2006). The anisotropy was always well below the whistler threshold. Note that these thresholds are based on the total temperature anisotropy, whereas the anisotropy we plot is the core anisotropy. Adding the suprathermals would move data even farther from the firehose thresholds. There was no radial dependence of the temperature anisotropy, as would be expected if the electrons were expanding adiabatically, indicative of significant scattering and/or energization. In addition, there was no radial dependence of the $T_{e\perp}/T_{e\parallel}$ vs $\beta_{e,\parallel}$ relationship, as also found by Štverák et al. (2008) for distances from .3 to 4 au.

4. Discussion

We have presented results of a statistical study examining the dependence of whistler-mode wave occurrence on distance from the Sun, and the association of wave occurrence with selected electron parameters. Waves were identified using an auto-identification code operating on the bandpass filter electric field and search coil magnetic field data from PSP encounters 1 through 9, covering solar radial distances from ~16 Rs to ~160 Rs ( ~.07 to ~.7 au). The most significant finding is that whistlers are extremely rare inside ~28 Rs (~.13 au). This has important implications for the evolution of strahl electrons and regulation of heat flux close to the Sun.

The most plausible explanation for the lack of whistlers inside ~28 Rs is that the other wave modes in similar frequency bands, including ion acoustic waves, other electrostatic waves and nonlinear waveforms including solitary waves, are more easily destabilized or that the electron distributions are not unstable to the whistler-mode. Studies of electron distributions close to the Sun have revealed a deficit, which increases closer to the Sun, in electrons propagating sunward, due to the ambipolar electric field (Berčič et al., 2021b; Halekas et al., 2021b). It is likely that the changes in the functional form of the electron distributions due to this deficit alter the modes that are destabilized (Halekas et al, 2021b; Berčič et al., 2021a). Halekas et al (2021b) found that, when the deficit was observed, the heat flux-$\beta_{e,||}$ relationship was not consistent with the whistler heat flux fan instability.

Linear stability analysis (e.g. Krafft and Volokitin, 2003, Vasko et al., 2019, Verscharen et al, 2019, López et al, 2020) and simulations (Roberg-Clark et al. 2016, 2018a,2018b; Komorov et al, 2018; Micera et al 2020) have been used to address the excitation of whistlers and other modes by heat flux instabilities and the resulting change in strahl and heat flux. A number of studies have explicitly focused on changes associated with radial expansion of the solar wind (Micera et al, 2021, Schroeder et al, 2021, Sun et al, 2021).

The instability threshold for whistlers may also be higher closer to the Sun. The whistlers are observed in a narrow range of $\beta_{e,||}$ from ~ 1 to 10, with normalized heat flux values consistent with excitation by the whistler heat flux fan instability. The linear stability analysis, performed for $\beta_{e,||}$ from 0.01 to 5, showed that $Q_{norm}$ must be larger at low $\beta_{e,||}$ (Vasko et al., 2019). Low beta (<.1) was only rarely observed in E1-E9, and only very close to Sun (<20 Rs) when $Q_{norm}$ was well below the threshold. Although $\beta_{e,||}$ is often <1, in the cases inside ~25 Rs, $Q_{norm}$ was also well below threshold. The few whistlers observed when $\beta_{e,||}$<1 (all were >.7) occurred at r>35 Rs. López et al. (2020) explored the different instabilities (parallel and oblique whistler heat flux instabilities, electron acoustic instability and electron beam instability) that could be excited for .1< beta <10, and the dependence of thresholds on beta and the ratio of the strahl speed to the electron Alfven speed, $V_s/V_{Ae}$. They concluded that more than one instability likely operated simultaneously and/or sequentially to reduce the heat flux. The dependence on $V_s/V_{Ae}$ was also addressed by Sauer and Sydora (2010), Verscharen et al (2019),and Micera et al (2021). As the magnetic field increases closer to the Sun, this ratio would decrease, causing the system to be less unstable to whistlers.

Using PIC simulations with an imposed heat flux focused on astrophysical settings, Roberg-Clark et al.(2018a) and Komorov et al. (2018) concluded that in a high beta plasma (~ 10 to ~100), the heat flux is controlled by oblique whistlers and scales as beta$^{-1}$. As shown in Figure 4f, when whistlers are observed (beta from ~1 to 10), this scaling is seen; however, at both high beta(>50) and low beta (<~.7), whistlers were not observed in the PSP data and the heat flux did not scale with beta (5g). In addition, close to the Sun (inside ~28 Rs) the scaling was not observed, as is also the case when the sunward electron deficit occurs (Halekas et al., 2021b).

Micera et al. (2020), initializing their PIC with electron distributions based on PSP data at ~.2 au and beta~2, found that initially oblique whistlers were excited, scattered the strahl, and later quasi-parallel whistlers grew and further scattered the electrons. In a follow-up study using an expanding box simulation, Micera et al. (2021) examined the radial evolution of the waves excited by an initially stable combination of strahl and core electrons. Oblique sunward propagating whistlers were first destabilized, at greater radial distances, both quasi-parallel and oblique waves occurred. In addition, firehose instabilities were excited. Which instabilities were excited depended on $V_s/V_{Ae}$, beta, and other wave modes present.

Wave amplitudes increase as the satellite gets closer to the Sun; wave electric field amplitudes reach 45 mV/m and the ratio of the wave magnetic field to the background field can reach ~0.1. Because these are values for a single component, the actual amplitudes are larger. The decrease in wave amplitudes with radial distance is consistent with earlier studies at distances >0.3 au. Beinroth and Neubauer (1981) found that the integrated energy in the whistler band decreased by a factor of ~30 from 0.3 au to 1 au. Lengyel-Frey et al. (1996) also showed that this decrease continued to at least ~3 au; at greater distances, the wave power would have been below background. The limited statistics from a comparison of waveform capture data from PSP Encounters 1 through 4 (outside ~.15 au) to captures from STEREO at 1 au Cattell et al (2021a) found that electric field amplitudes decreased with distance from the Sun, but average values at ~.3 au were comparable to those at 1 au. All these studies also found a decrease in wave frequency with radial distance, consistent with the decrease in the magnetic field.

The most important parameters determining the strength of the interaction of electrons with whistler-mode waves are the wave amplitude and the wave angle (Vo et al, 2021). Figure 5 plots the wave amplitude as a function of wave propagation angle for the events identified in encounter 1 (when all three components of the wave magnetic field were accurately measured so that the wave angle can be determined via the singular value decomposition method of Santolik et al., 2003). Most of the whistlers during encounter 1 were quasi-parallel. There is no dependence of average amplitude on wave **propagation** angle **with respect to the solar wind magnetic field**, in contrast to 1 au where the largest amplitude whistlers are oblique (Vo et al., 2021). The increase of wave amplitudes close to the Sun (shown in Figure 3), in combination with particle tracing results, indicates that the whistlers at these radial distance (outside 28 Rs) are very effective at scattering electrons over a range of energies.

5. Conclusions

We have presented results of a statistical study to examine the dependence of whistler-mode wave occurrence on distance from the Sun, and the association of wave occurrence with selected electron parameters. Waves were identified using an auto-identification code operating on the bandpass filter electric field and search coil magnetic field data from PSP encounters 1 through 9, covering radial distances from ~16 Rs to ~160 Rs (~.07 to ~.7 au). The electron parameters were examined for times when waves were observed within 15 seconds of the center time of the electron measurement and compared to times when wave data were obtained but no whistlers were identified within 15 s of the electron measurement. We found that:

    1. Almost no narrowband whistler mode waves were observed inside ~28 Rs (~.13 au);

    2. Whistlers occurred in a narrow range of parallel electron beta, from ~1 to 10;

    3. When whistlers were observed, the normalized heat flux ($Q_{norm}$)-parallel electron beta ($\beta_{e,||}$) relationship was constrained by the heat flux fan instability threshold;

    4. The average whistler electric field amplitudes increased as distance from the Sun decreased, but the normalized magnetic field magnitudes were approximately constant;

    5. The average whistler (spacecraft frame) frequencies normalized to the electron cyclotron frequency were ~.1 to ~.2;

    6. During Encounter 1 whistlers were observed with wave propagation angles from quasi-parallel to highly oblique, and there was no dependence of wave amplitude on wave propagation angle;

    7. Inside ~28 Rs, large amplitude narrowband electrostatic waves were observed in the same frequency range as the narrowband whistlers($f_{lh}$ to ~.5 $f_{ce}$);

8. Close to the Sun, $Q_{norm}$ decreased (as also observed by Halekas et al., 2021 for all electron measurements for E1, E2, E4 and E5), and the $Q_{norm}$-$\beta_{e,\|}$ relationship was not constrained by the heat flux fan instability; and,

    9. For low $\beta_{e,\|}$ (<~.1), which occurred primarily inside ~25 Rs, $Q_{norm}$ was below heat flux instability thresholds. For high $\beta_{e,\|}$ (~>50), $Q_{norm}$ was above heat flux instability thresholds, and did not depend on $\beta_{e,\|}$, indicating that heat flux instabilities were not operating.

The transition from whistler-mode waves to electrostatic waves occurred over similar radial distances as the sunward deficit in the electrons due to the ambipolar electric field (Halekas et al., 2021b, Berčič et al., 2021b), and the increase in the relative density of halo electrons (Halekas et al., 2021b). These facts support the idea that the ambipolar electric field results in changes in the form of the electron distributions that are more unstable to the electrostatic waves than to whistlers. Outside ~30 Rs, there is strong evidence that the narrowband whistlers scatter strahl to produce the halo and regulate heat flux.

The surprising observation that whistler-mode waves are almost never observed inside ~28 $R_s$ (~0.13 au) is crucial for understanding the evolution of the solar wind close to the Sun. The parallel electron beta and normalized heat flux in these radial distances often have values comparable to those farther out. Changes in the form of the electron distribution associated with the ambipolar electric field or changes in other plasma properties must result in lower instability limits for the other modes (including the observed solitary waves, ion acoustic waves) that are observed. The lack of narrowband whistler-mode waves close to the Sun and in regions of either low (<.1) or high (>~10) beta may also be significant for the understanding and modeling of the evolution of flare-accelerated electrons, other stellar winds, the interstellar medium, and intra-galaxy cluster medium.

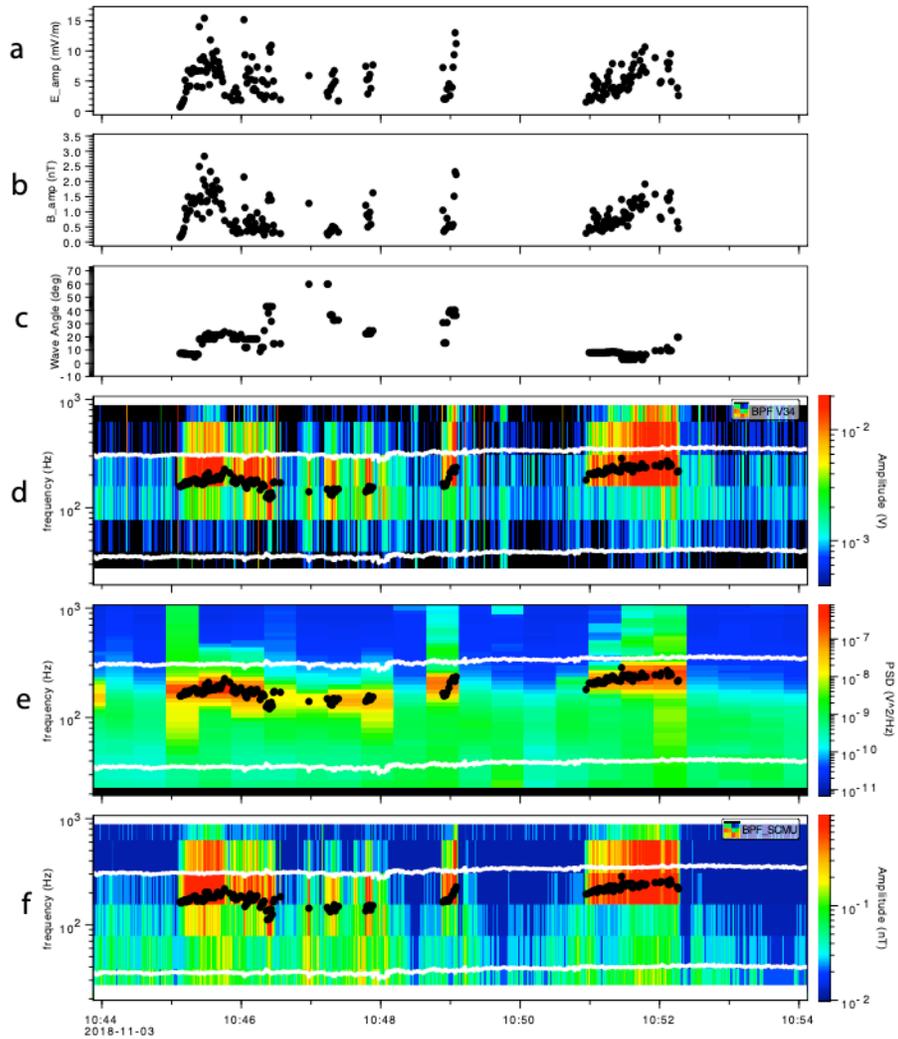

Figure 1. Comparison of whistlers auto-identified and the original electric and magnetic field bandpass filter data for 10 minutes on 03 November 2018. a. electric field amplitude, b. BPF wave magnetic field amplitude, c. wave propagation angle, identified whistler wave event frequency over-plotted in black on the d. BPF electric field, e. electric field spectrum, and f. BPF magnetic field. White lines represent $f_{lh}$ and $0.2\ f_{ce}$.

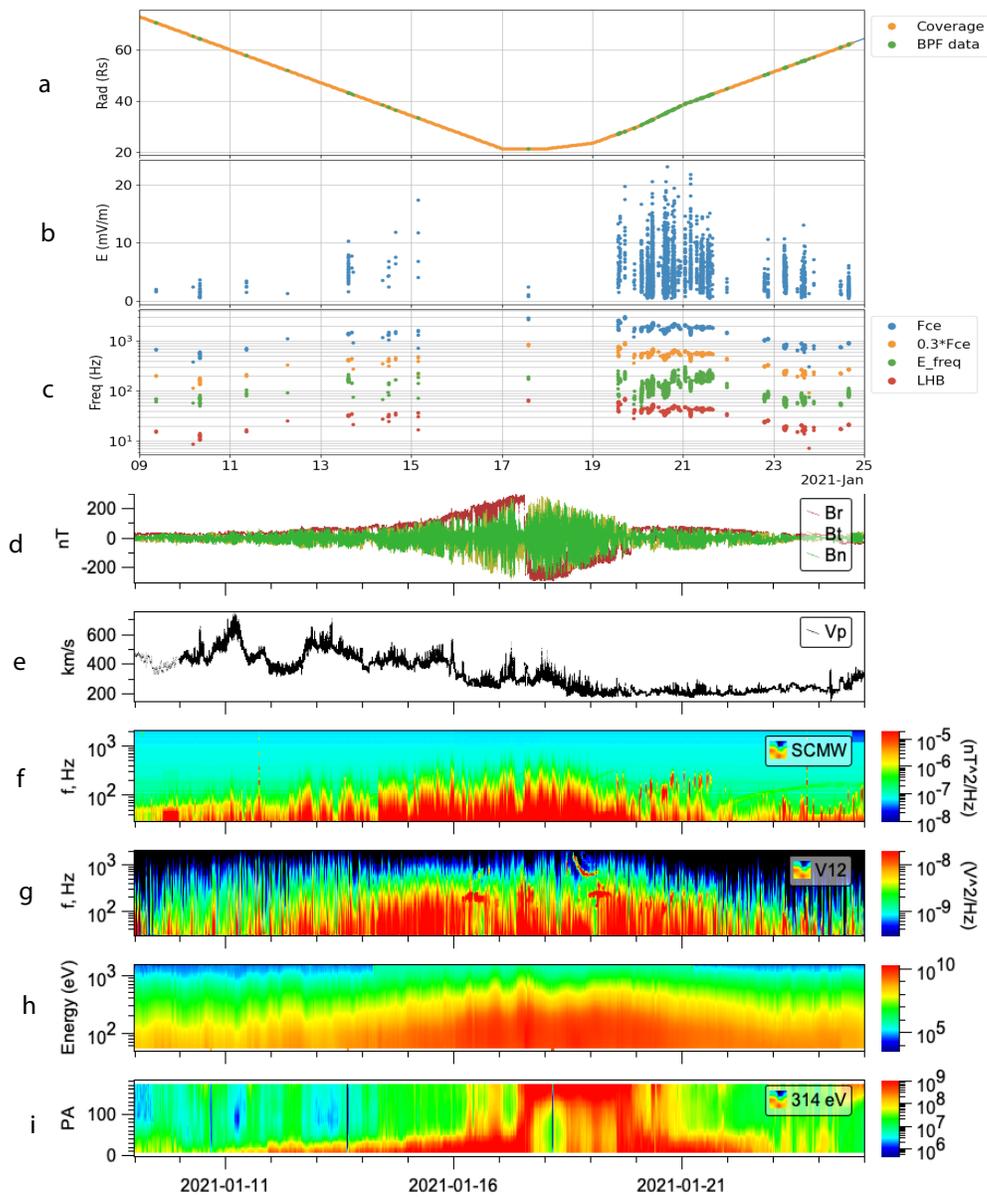

Figure 2. Encounter 7 comparison of whistler identification. From top to bottom: the radial distance in AU (orange dots indicate BPF data, green identified as whistlers); wave electric field amplitude; relevant frequencies, the wave frequency in green, the electron cyclotron frequency, $f_{ce}$, in blue, 0.3 $f_{ce}$ in yellow, and the lower hybrid frequency, $f_{lh}$, in red; magnetic field in RTN coordinates, proton speed from SPAN-I, DC-coupled spectra of magnetic field and electric field 30-5000 Hz, electron energy spectrum and 314 eV pitch angle spectrum.

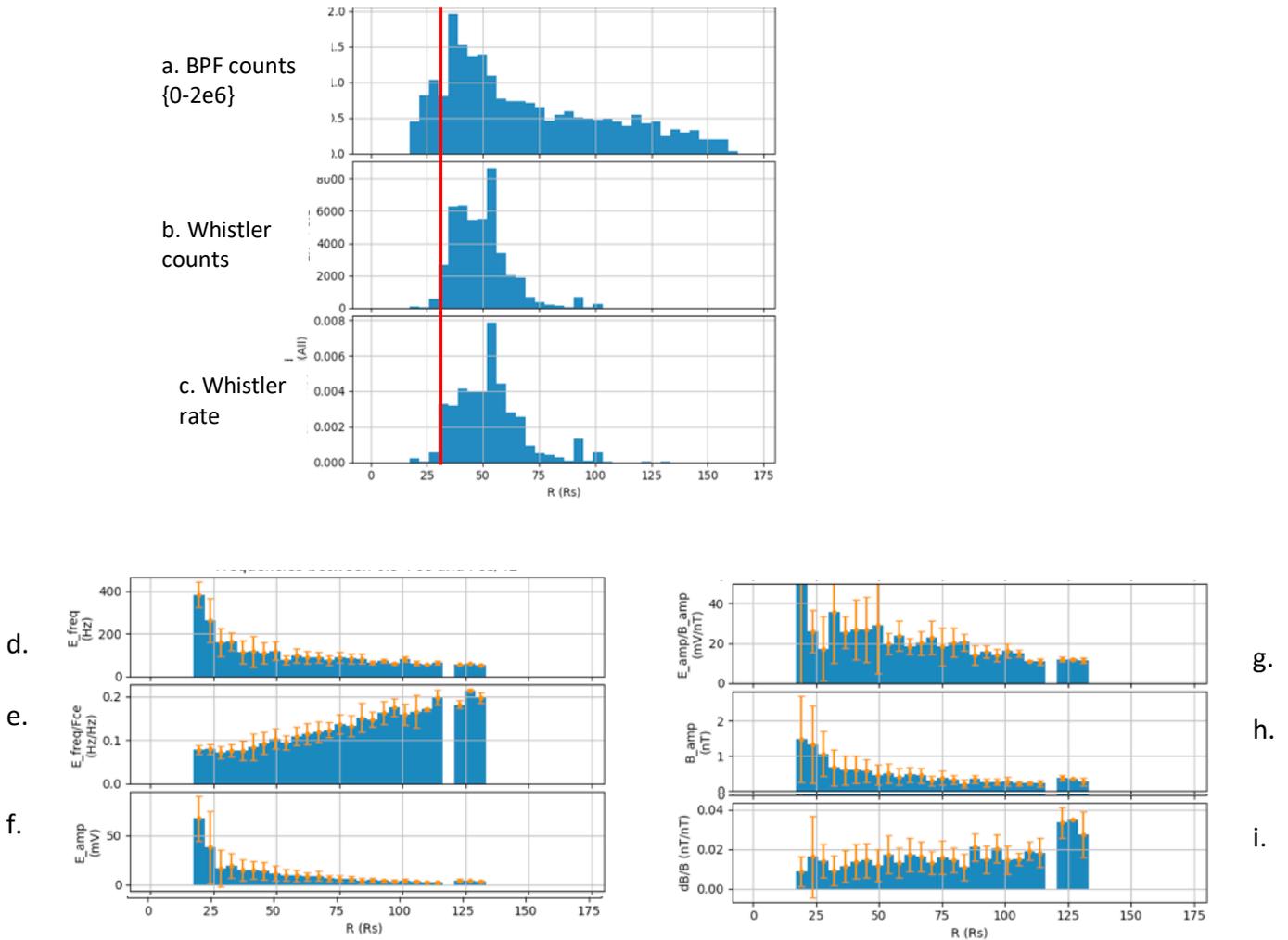

Figure 3 Statistics of whistler-mode waves identified in bandpass filter data. (a) Number of BBF samples in E1 through E9 versus radial distance (b) number of BBF samples identified as whistler-mode wave versus radial distances; and (c) whistler-mode wave occurrence rate .versus radial distance. The red line shows location where whistler occurrence drops (~28 $R_s$ or ~.13 au). (d) Spacecraft frame frequency, f; (e) f/fce; (f) Electric field amplitude, dE, mV/m; (g) dE/dB; (h) magnetic field amplitude, dB, nT; and (i) ratio of wave magnetic field to background magnetic field, dB/B.

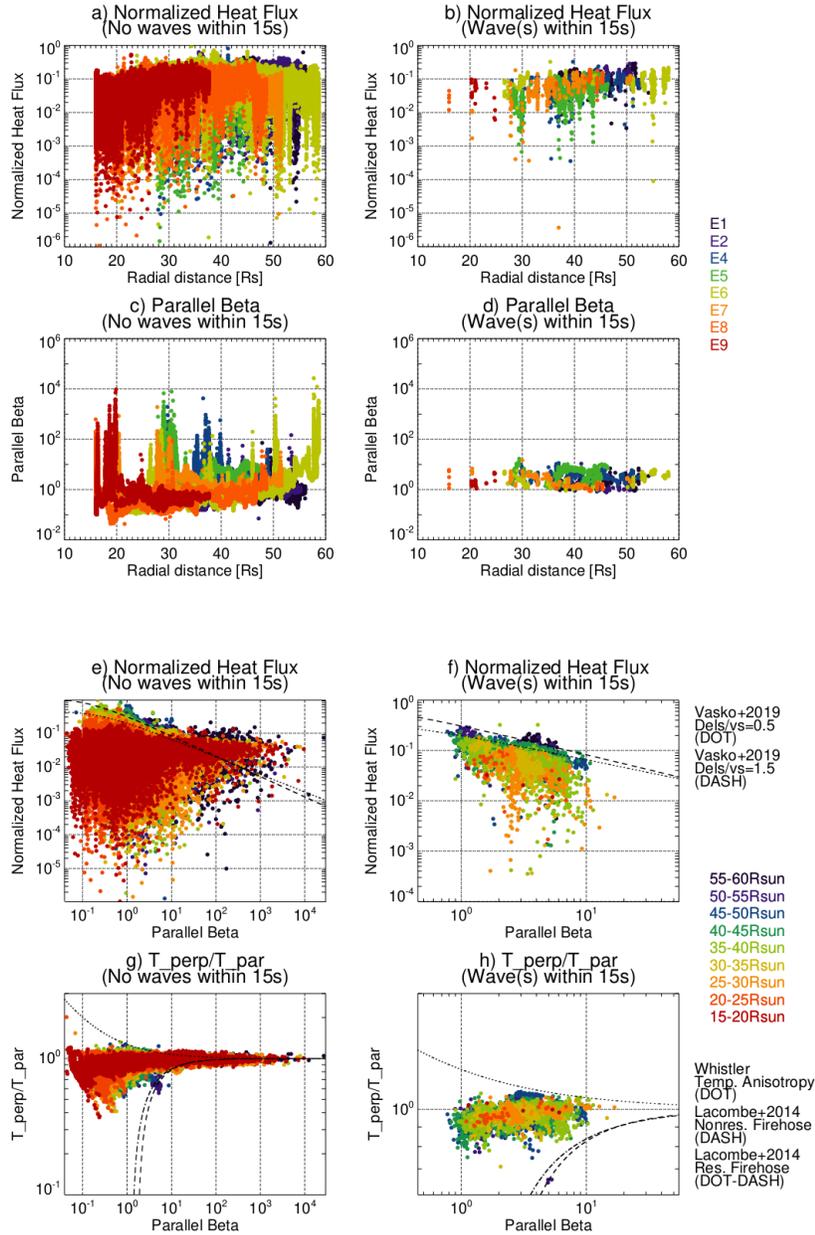

Figure 4. Comparison of electron parameters when whistlers were not identified within 15 seconds of the electron measurement(left) and when at least one whistler identified within 15 seconds (right). (a) and (b) the normalized heat flux, $Q_{norm}$, and (c) and (d) $\beta_{e,||}$, plotted versus radial distance and color coded by encounter number. (e)-(h) are plotted versus $\beta_{e,||}$, and color coded by radial distance. (e) and (f) $Q_{norm}$ vs $\beta_{e,||}$) with the Vasko et al (2019) heat flux fan instability threshold over-plotted; (g) and (h) core temperature anisotropy vs $\beta_{e,||}$ with the temperature anisotropy and firehose instability limits over-plotted. The upper dotted line is the temperature anisotropy threshold, and the lower lines are arbitrary firehose resonant (dot-dash). and non-resonant (dash) instability ( Lacombe et al. 2014, based on Gary et al., 2006). Note that the right hand plots in (c) and (d) are for a restricted range of beta.

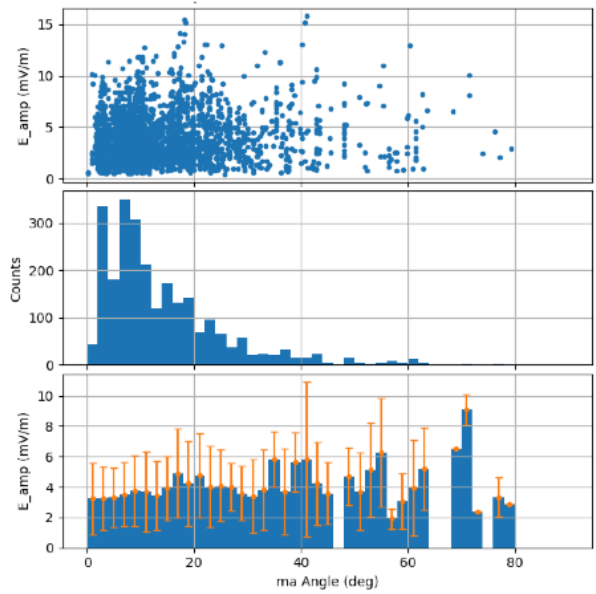

Figure 5. Wave amplitude versus wave propagation angle for Encounter 1 whistlers. From top to bottom: Scatter plot of wave amplitude versus angle; histogram of number of whistlers at a given angle; and average amplitude versus angle.

Acknowledgements: We acknowledge the NASA Parker Solar Probe Mission, the FIELDS team led S. D. Bale, and the SWEAP team led by J. Kasper for use of data. The FIELDS and SWEAP experiments on the Parker Solar Probe spacecraft were designed and developed under NASA contract NNN06AA01C. Data analysis was supported under the same contract. TD acknowledges support from CNES. We thank the developers of the scipy (scipy.org), matplotlib(matplotlib.org) and pypi (pypi.org) software used in analysis of the wave data.

Appendix

The auto-identification code uses the dc-coupled bandpass filter data obtained in 13 frequency bands for one electric field and one search coil magnetic field channel. The code uses a non-linear fitting approach to identify BPF samples consistent with a narrowband wave and to determine their frequency and amplitude, by minimizing the error between the measured BPF values and the expected response of the BPF to a sine wave with this frequency and amplitude. The frequency that minimizes the error was first determined. Next the gain at that frequency was used to correct the measured peak value to determine the amplitude of the narrowband wave. We used five frequency bins centered around the bin with the maximum peak to fit to a sine wave response. The standard deviation of the residuals was used the error metric. The minimum error also provides a quality metric for the fit. For this study, we used a minimum fit error of less than 0.1, a limit that was checked in two ways. The time domain waveform measurements previously identified as narrowband whistler-mode waves were run through a simulated filter bank, and the artificial data were then run through our auto-id code. The results were compared to the original waveform to verify the peak amplitude and frequency of the waveform were calculated correctly. Note that a similar approach using various artificial waveforms was used to develop the code. In addition, for a number of different intervals, we compared times identified as containing whistlers by the auto-id code with those identified by eye.

Figure A1 provides an overview of some properties obtained by the auto-identification code for Encounters 1 through 9. Panel a plots the PSP orbit versus radial distance in AU with the coverage of the BPF data as orange dots and identified whistlers as green dots. The wave electric field amplitudes for identified whistlers are shown in panel b. Panel c plots the wave frequency in green; the electron cyclotron frequency, $f_{ce}$, in blue, 0.3 $f_{ce}$ in yellow, and the lower hybrid frequency, $f_{lh}$, in red are also plotted. Panel d plots the ratio of the wave frequency identified in the magnetic field, $f_B$, to that that identified in the electric field, $f_E$. Panel e plots the ratio of $f_E/f_{ce}$. Although the time-scale of this figure is too long to see details, there are several features that can already be discerned. Wave amplitudes increase and there are fewer whistlers as PSP gets closer to the Sun. In addition, there are some encounters with an asymmetry between the number of identified whistlers between the inbound and outbound orbits when the sampling is fairly symmetric (E4, E7, and E8); the possible causes are under investigation.

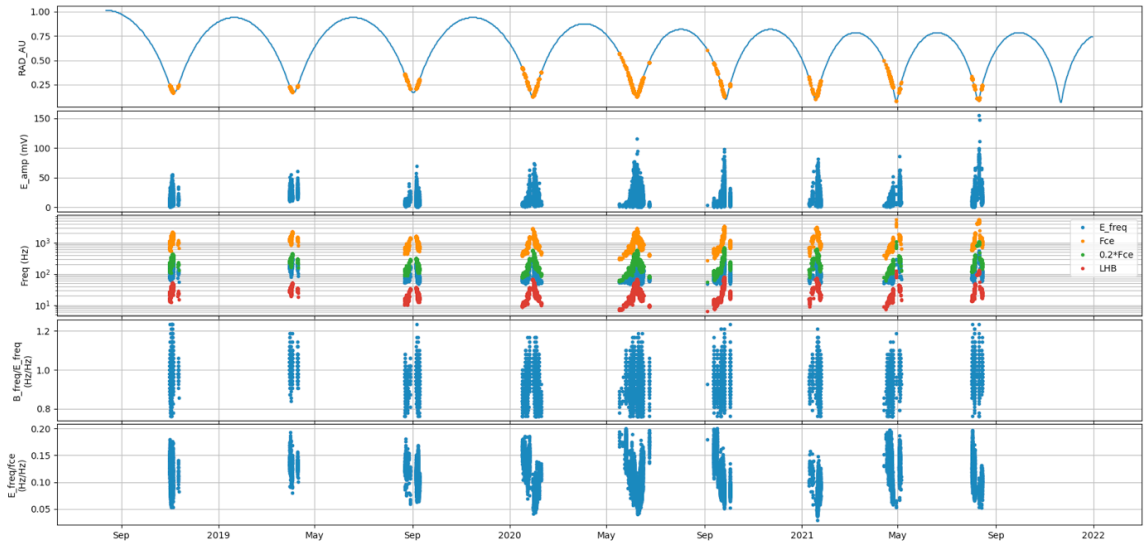

Figure A1. Location of identified whistler-mode waves in Encounters 1-9. From top to bottom: Radial distance in AU (orange dots indicate BPF data, green identified as whistlers), wave electric field amplitude, relevant frequencies, $f_{BW}/f_{EW}$, and $f_{EW}/f_{ce}$